\documentclass[12pt,letterpaper]{article}
\pdfoutput=1
\usepackage{jheppub}
\usepackage{amssymb}
\usepackage{color}
\usepackage{slashed}
\usepackage{cancel}
\usepackage{amsmath}
\usepackage{amsfonts}
\usepackage{amssymb}
\usepackage{graphicx,subcaption}
\usepackage{caption}
\usepackage{subcaption}
\usepackage{float}
\usepackage{tikz}
\usetikzlibrary[arrows]
\usepackage{epsfig}
\usepackage{appendix}
\usepackage{listings}
\usepackage{xcolor}
\usepackage{mathabx}

\newcommand{\be}{\begin{equation}}
\newcommand{\ee}{\end{equation}}

\newcommand{\highlight}[1]{\colorbox{lightgray}{$\displaystyle #1$}}

\title{Schwinger-Dyson approximants}

\author{Bartomeu Fiol,}
\author{Elena Gij\'on}
\author{and Unai Lejarza Alonso}

\affiliation{Departament de F{\'\i}sica Qu\`antica i Astrof\'isica i \\Institut de Ci{\`e}ncies del Cosmos, 
Universitat de Barcelona,
Mart{\'\i}\ i Franqu{\`e}s 1, 08028 Barcelona, Catalonia, Spain}

\emailAdd{bfiol@ub.edu}
\emailAdd{egijonru22@alumnes.ub.edu}
\emailAdd{ulejaral82@alumnes.ub.edu}

\abstract{We revisit the solution to the Schwinger-Dyson equations in the simple case of the 0-dimensional $\frac{1}{2}m^2 \phi^2 +\frac{\lambda}{4} \phi^4$ theory with $m^2>0$ and $\lambda \geq 0$. We argue that the truncated Schwinger-Dyson equations are solved by rational approximants to all n-point functions $\langle \phi^{2k} \rangle$, and provide strikingly simple recursive relations for them. These rational approximants are constructed without any reference to ordinary perturbative expansions. They turn out to be Pad\'e approximants for $\langle \phi^2 \rangle$ and for half of the truncations in the case of $\langle \phi^4 \rangle$, but they are not Pad\'e approximants for higher n-point functions. This difference is related to the fact that $\langle \phi^2 \rangle$ and $\langle \phi^4 \rangle$ are Stieltjes functions, while higher n-point functions are not. We prove that as the size of the truncation tends to infinity, these rational approximants converge to the full non-perturbative n-point functions for all positive values of the coupling $\lambda$. Thus, in the example studied in this work, these new rational approximants are much easier to derive than the usual Pad\'e approximants, and when different, they are better suited to approximate the full non-perturbative n-point functions.}

\begin{document}
\maketitle
\section{Introduction}

There is more to Quantum Field Theory (QFT) than perturbative expansions. It has been known since the early days of QFT that generic perturbative expansions are asymptotic \cite{Dyson:1952tj, LeGuillou:1990nq}, so they may miss qualitatively important features of the theory. This has motivated the development of a variety of non-perturbative approaches to study QFT. Among those approaches, one of the earliest ones are the Schwinger-Dyson (SD) equations \cite{Dyson:1949ha, Schwinger:1951ex, Schwinger:1951hq} : these are an infinite set of equations relating the n-point functions of the theory. For physically relevant four-dimensional gauge QFTs like QED or QCD,  the full set of SD equations is impossibly hard to solve. One then resorts to various truncation schemes, and has to deal with issues related to gauge-fixing and renormalization, see  \cite{Mandelstam:1979xd, Cornwall:1981zr, Cornwall:1989gv, Curtis:1990zs, Roberts:1994dr, Alkofer:2000wg} for a necessarily incomplete list of references.

To explore truncation schemes for the Schwinger-Dyson equations, a possible strategy is to test different approaches on
simpler, lower dimensional QFTs where exact results are known, see {\it e.g.} \cite{Bender:1976xh, Dorey:1989mr, Adam:1994by}. Taking this philosophy to the extreme, there has been plenty of work applying the Schwinger-Dyson equations to 0-dimensional theories, where path integrals reduce to ordinary integrals \cite{Bender:1988bp, Okopinska:1990pt, Garcia:1996np, Bender:2022eze, Bender:2023ttu, Peng:2024azv, Banks:2024ydh}. While in these toy models the thorny issues of gauge-fixing or renormalization are simply absent, their solutions already display non-analytic behavior, which makes them an interesting training ground. In this work, we join this line of research and revisit the familiar case of the zero-dimensional  Euclidean $\lambda^4$ theory,
\be
S=\frac{1}{2}m^2 \phi^2 +\frac{\lambda}{4}\phi^4
\label{theaction}
\ee
in the case $m^2>0$ and $\lambda \geq 0$. The aim of this work is to use the Schwinger-Dyson equations to obtain approximants to the exact n-point functions of the theory for the choice of integration contour along the real axis,
\be
\langle \phi^{2k} \rangle = \frac{ \int_{-\infty}^\infty d\phi \,\, \phi^{2k} \,\, e^{-S(\phi)}} { \int_{-\infty}^\infty d\phi \,\, e^{-S(\phi)}}
\ee
and to demonstrate the convergence of these new approximants to the exact non-perturbative n-point functions, for all positive values of the coupling $\lambda$.

We will argue that for the theory at hand, the truncation of the infinite tower of SD equations to the first $n$ SD equations is solved by rational approximants to the 2k-point functions, that we call the Schwinger Dyson approximants, and denote by $\langle \phi^{2k} \rangle^n_{\text{SD}}$. These rational approximants are not entirely new: to the best of our knowledge, they first appeared in \cite{Bender:1988bp}, where it was claimed that their validity is limited to the weak coupling region, a claim we disagree with. Much more recently, these rational approximants appeared again in \cite{Konosu:2024zrq}, who arrived at them from a superficially different starting point: the homotopy algebra approach to QFT \cite{Hohm:2017pnh, Jurco:2018sby}. The arguments in \cite{Bender:1988bp} and the present work show that the main result of \cite{Konosu:2024zrq} can be derived immediately from the SD equations, without having to rely on the homotopy algebra framework\footnote{See \cite{Okawa:2022sjf, Konosu:2023psc, Konosu:2024dpo} for proofs that the correlation functions obtained with the homotopy algebra approach to QFT satisfy the SD equations.}. On the other hand, the authors of \cite{Konosu:2024zrq} claim but do not prove that these rational approximants capture also the strong coupling regime of the exact n-point functions. Making precise this claim was the original motivation for the present work, and our two main results are an explicit recursive definition of the novel rational approximants, and the proof of their convergence to the exact n-point functions, for all positive values of the coupling.

The structure of the paper is as follows. In section \ref{0dimensions} we present the exact n-point functions $\langle \phi^{2k} \rangle$ and recall the well-known power series, $\langle \phi^{2k} \rangle_{\text{weak}}$ and $\langle \phi^{2k} \rangle_{\text{strong}}$. We then discuss the analytic properties of the exact $\langle \phi^{2k} \rangle$ and point out a difference that will play a major role in the rest of the paper: as functions of $\lambda$ in the complex plane $\langle \phi^{2} \rangle (\lambda)$, $\langle \phi^{4} \rangle (\lambda)$ are Stieltjes functions\footnote{The definition and basic facts about Stieltjes functions are reviewed in appendix \ref{stieltjesapp}.}, while higher n-point functions are not Stieltjes. Nevertheless, we show that $
\frac{\langle \phi^{2k+2} \rangle}{\langle \phi^{2k} \rangle}$ are Stieltjes functions for $k\geq 0$. This has the important implication that while $\langle \phi^6 \rangle$ are higher n-point functions are not Stieltjes, they are equal to the product of a finite number of Stieltjes functions,
\be
\langle \phi^{2k} \rangle = \frac{\langle \phi^{2k} \rangle}{\langle \phi^{2k-2} \rangle} \frac{\langle \phi^{2k-2} \rangle}{\langle \phi^{2k-4} \rangle}\dots \frac{\langle \phi^{4} \rangle}{\langle \phi^{2} \rangle} {\langle \phi^2 \rangle}.
\label{prodstieltjesintro}
\ee
This observation will provide a useful perspective on the rational approximants we discuss next.

We then turn to the main topic of this work, and solve the truncated Schwinger-Dyson equations of the theory (\ref{theaction}) by rational approximants defined recursively. Namely, introducing the families of recursive polynomials in $\lambda$ labelled by $k=0,1,2,\dots$,
\be
p_n^{(k)}(\lambda)=
\begin{cases}
0 & n<k+1 \\
1 & n=k+1 \\
p_{n-1}^{(k)}(\lambda)+(2n-3) \lambda p_{n-2}^{(k)} & n > k+1
\end{cases}
\ee
we  define the rational approximants
\be
\langle \phi^{2k} \rangle_{\text{SD}}^n= (2k-1)!! \, \frac{p_n^{(k)}(\lambda)}{p_{n}^{(0)}(\lambda)}
\label{sdapproxintro}
\ee
which satisfy the first $n-1$ Schwinger-Dyson equations. As discussed in the main text, these rational approximants have a number of appealing properties, {\it e.g.} for a given truncation, approximants to different 2k-point functions all have the same denominator, and they don't present spurious zeros or poles on the positive real axis.

In section \ref{comparison} we proceed to compare the new rational approximants $\langle \phi^{2k} \rangle_{\text{SD}}^n$ to the most familar rational approximants, the Pad\'e approximants obtained from asymptotic weak coupling series $\langle \phi^{2k}\rangle_{\text{weak}}$. We rigorously prove that for the 2-point function, $\langle \phi^{2} \rangle_{\text{SD}}^n$ are precisely the familiar [N,N] and [N,N+1] Pad\'e approximants. For the 4-point function, half of the new approximants $\langle \phi^4 \rangle_{\text{SD}}^n$ are precisely the [N,N+1] Pad\'e approximants, while the other half don't coincide with any Pad\'e approximant. Notably, starting with $\langle \phi^6 \rangle_{\text{SD}}^n$, the new approximants do not coincide with any Pad\'e approximants to $\langle \phi^{2k}\rangle_{\text{weak}}$. Rather, they are products of Pad\'e approximants to the Sieltjes functions in (\ref{prodstieltjesintro}). This provides a compelling perspective on the usefulness
of the SD approximants: if a function is Stieltjes, Pad\'e approximants are the most convenient rational approximants \cite{baker, Bender:1999box}. For functions that are not Stieltjes, conventional Pad\'e approximants do not necessarily possess the same nice properties, {\it e.g.} as we will see explicitly in the main text, they can display spurious zeros or poles. If the function is not Stieltjes but a product of Stieltjes functions as in (\ref{prodstieltjesintro}), an alternative to the Pad\'e approximants could be to derive Pad\'e approximants to each of the Stieltjes functions, multiply them, and obtain a rational approximant to the non-Stieltjes function. In the case at hand, the SD approximants to higher n-point functions directly provide the end result, avoiding the process outlined above.
 

In section \ref{convergence}, we prove that for any $\langle \phi^{2k} \rangle$, the rational approximants (\ref{sdapproxintro}) converge to the exact $\langle \phi^{2k} \rangle$ as $n\to \infty$. We present two arguments, with a common key ingredient: we prove that $\frac{\langle \phi^{2k+2} \rangle}{\langle \phi^{2k} \rangle}$ is the unique Stieltjes function that has $\frac{\langle \phi^{2k+2} \rangle_{\text{weak}}}{\langle \phi^{2k} \rangle_{\text{weak}}}$ as Stieltjes series. This fact is extremely helpful since there exists a well-developed theory for the convergence of the relevant Pad\'e approximants to Stieltjes series \cite{baker, Bender:1999box}. Armed with this result, we provide two one-line proofs of the convergence of  $\langle \phi^{2k} \rangle_{\text{SD}}^n$ to the exact $\langle \phi^{2k} \rangle$. For the sake of comparison, for higher n-point functions we are not aware of any rigorous argument that would prove the convergence of the usual  Pad\'e approximants to the exact non-perturbative answer.

The main text is complemented with a number of appendices. The first one collects various basic facts about Stieltjes functions, while the other three are devoted to technical details and alternative proofs to some statements in the main text.

Let's conclude the introduction by first summing up what we think are the main lessons of this work, and suggesting directions for further research. For the theory under study, the 2-point function is a Stieltjes function of the coupling, and the novel approximants are just the [N,N] and [N,N+1] Pad\'e approximants; the only advantage in this case is that  our derivation bypasses the construction of the asymptotic power series $\langle \phi^2 \rangle_{\text{weak}}$. On the other hand, for higher point functions, starting with $\langle \phi^6 \rangle$ and above, we advocate that instead of trying to approximate them by the familiar Pad\'e approximants, the rational approximants that solve the truncated SD equations are better approximants. Moreover, at least for this theory, these novel approximants can be thought of as products of Pad\'e approximants to Stieltjes functions. 

Still in the realm of 0d toy models of QFTs, one can consider other potentials like $V(\phi)=\phi^{2m}$ or $V(\phi)=i\phi^3$ which is a toy model for PT-symmetric QFTs \cite{Bender:1999ek, Bender:2005tb, Bender:2007nj}. The SD equations are algebraic for these toy models, so we expect that the SD approximants will be rational functions, and it will be interesting to elucidate in which cases they are products of Pad\'e approximants.

In higher dimensions, the SD equations are no longer algebraic, so solving even a truncated set of them becomes much more complicated. One possible venue of research suggested by the present work is to take a closer look at what quantities are Stieltjes functions, since their products are under as good control as Stieltjes functions. In generic QFTs, the K\"all\'en-Lehmann spectral representation of exact 2-point functions in momentum space involves a positive definite spectral function, so under some conditions these are Stieltjes functions \cite{Peris:2006ds, Masjuan:2009wy}. Also, in some simple theories, it is known that energy eigenvalues are Stieltjes functions of the coupling \cite{Loeffel:1969rdm}. For theories where this is not the case, it is worth exploring if there are combinations of eigenvalues  - {\it e.g.} their quotients - that are Stieltjes, as their approximants would be under better control.

\section{$\lambda \phi^4$ in 0 dimensions}
\label{0dimensions}
In this work we are going to focus on a single example, the Euclidean 0-dimensional $\phi^4$ theory \cite{Zinn-Justin:1979jnt}. The action is 
\be
S=\frac{1}{2}m^2 \phi^2 +\frac{\lambda}{4}\phi^4
\ee
and we will consider the case $m^2>0$ and $\lambda \geq 0$. Note that we follow the convention $\frac{\lambda}{4}$ for the coupling, rather than the more common $\frac{\lambda}{4!}$, since the expressions that we will derive are simpler in this convention. For this action, the partition function can be defined over three homologically independent contour integrals over the complex $\phi$ plane \cite{Bender:1988bp, Garcia:1996np, Witten:2010zr, Aniceto:2018bis}. We make the choice of integrating along the real axis, so the Euclidean partition function is
\be
Z(j)=\frac{1}{\sqrt{2\pi}} \int_{-\infty}^\infty d\phi \,\, e^{-S(\phi)+j\phi}
\label{partfunction}
\ee
For this choice of contour the odd n-point functions vanish, and we are left with the even ones,
\be
\langle \phi^{2k} \rangle = \frac{ \int_{-\infty}^\infty d\phi \,\, \phi^{2k} \,\, e^{-S(\phi)}} { \int_{-\infty}^\infty d\phi \,\, e^{-S(\phi)}}
\label{exactphi2k}
\ee
These exact n-point functions (\ref{exactphi2k}) satisfy the Schwinger-Dyson equations, that in this case boil down to a three-term recurrence relation
\be 
\lambda \langle \phi^{2k+2}\rangle =- m^2\langle \phi^{2k}\rangle+(2k-1) \langle \phi^{2k-2}\rangle
\label{sdeq}
\ee
for $k\geq 1$. The most important property of these relations is that they are linear in the n-point functions. One can also write Schwinger-Dyson equations for connected or 1PI correlation functions, but the resulting relations are not linear in the correlation functions \cite{Bender:2022eze, Bender:2023ttu,Peng:2024azv, Banks:2024ydh}. This linearity is the fundamental reason why we choose to work with ordinary n-point functions $\langle \phi^{2k} \rangle$.

Since - up to an overall factor of $\frac{1}{m^{2k}}$ - $\langle \phi^{2k} \rangle$ are functions of $\frac{\lambda}{m^4}$, from now on we set $m^2=1$. Owing to the extreme simplicity of this toy model, it is possible to obtain exact expressions for the exact n-point functions, in terms of parabolic cylinder functions $U_n(x)$ \cite{Garcia:1996np},
\be
\langle \phi^{2k} \rangle  =  \frac{(2k-1)!!}{(2\lambda)^{\frac{k}{2}}} \frac{U_k\left(\frac{1}{\sqrt{2\lambda}}\right)}{U_0\left(\frac{1}{\sqrt{2\lambda}}\right)}
\label{exactparabolic}
\ee
Using the recurrence relation for parabolic cylinder functions,
\be
U_{n+2}(x)=\frac{2}{2n+3} \left(U_n(x)-x U_{n+1}(x)\right)
\ee
it is immediate to check that (\ref{exactparabolic}) satisfy the SD equations (\ref{sdeq}). For some of the questions that we will address in the next sections, we find it more convenient to write the exact n-point functions (\ref{exactparabolic}) in terms of modified Bessel functions $K_\nu$. Using eqs. (12.7.10) and (12.7.11) of \cite{nistpara} we find
\be
\langle \phi^2 \rangle = 
 \frac{1}{2\lambda} \frac{K_{\frac{3}{4}}\left(\frac{1}{8\lambda}\right)-K_{\frac{1}{4}}\left(\frac{1}{8\lambda}\right)}{K_{\frac{1}{4}}\left(\frac{1}{8\lambda}\right)}
\label{exactphi2}
\ee
and in general
\be
\langle \phi^{2k} \rangle = 
\frac{(-1)^k}{2 \lambda^k} \frac{m_k(\lambda) K_{\frac{1}{4}}\left(\frac{1}{8\lambda}\right)-n_k(\lambda) K_{\frac{3}{4}}\left(\frac{1}{8\lambda}\right)}{K_{\frac{1}{4}} \left(\frac{1}{8\lambda}\right)} 
\label{exact2k}
\ee
where the polynomials $m_k(\lambda),n_k(\lambda)$ are defined recursively,
\be
\begin{split}
m_0=2,\hspace{.5cm} m_1=1, & \hspace{1cm} m_{k+1}(\lambda)=m_k(\lambda)+(2k-1)\lambda m_{k-1}(\lambda), \hspace{.5cm} k\geq 1 \\
n_0=0, \hspace{.5cm} n_1=1, &  \hspace{1cm} n_{k+1}(\lambda)=n_k(\lambda)+(2k-1)\lambda n_{k-1}(\lambda),
\hspace{.5cm} k\geq 1
\end{split}
\ee
In higher-dimensional theories, exact results are extremely hard to come by, and one typically resorts to perturbative approaches. In the case at hand, the perturbative series around $\lambda=0$ are obtained by expanding the $e^{-\frac{\lambda}{4}\phi^4}$ both in the numerator and denominator of (\ref{exactphi2k}), and exchanging the infinite sums and integrals. This exchange of infinite sums and integrals is not justified, as in the large $\phi$ regime, the $\phi^4$ term dominates over the $\phi^2$ one. If one nevertheless proceeds, one obtains perturbative series in $\frac{\lambda}{m^4}$,
\be
\langle \phi^{2k} \rangle_{\text{weak}} =
\frac{\sum_{n=0}^\infty \frac{1}{n!} (4n+2k-1)!! \left(-\frac{\lambda}{4 m^4}\right)^n}{\sum_{n=0}^\infty \frac{1}{n!} (4n-1)!! \left(-\frac{\lambda}{4 m^4}\right)^n}
\label{weak}
\ee
These series are asymptotic, as their radius of convergence is zero. Their usefulness is thus limited to small coupling. While these series are asymptotic, they formally satisfy the Schwinger-Dyson equations, eq. (\ref{sdeq}).  

A second possibility is to expand the $e^{-\frac{1}{2}m^2 \phi^2}$ terms in (\ref{exactphi2k}), and again exchange the infinite sums and integrals. This exchange is now justified, and one obtains strong coupling expansions in $\frac{m^2}{\sqrt{\lambda}}$,
\be
\langle \phi^{2k} \rangle_{\text{strong}} = \left(\frac{4}{\lambda}\right)^{\frac{k}{2}}
\frac{\sum_{n=0}^\infty \frac{1}{n!} \Gamma\left(\frac{2n+2k+1}{4}\right) \left(\frac{-m^2}{\sqrt{\lambda} }\right)^n }{\sum_{n=0}^\infty \frac{1}{n!}\Gamma\left(\frac{2n+1}{4}\right) \left(\frac{-m^2}{\sqrt{\lambda} }\right)^n }
\label{strong}
\ee
In contrast to the weak coupling perturbative expansions (\ref{weak}), these are convergent series. It is immediate to check that (\ref{strong}) satisfy the Schwinger-Dyson equations (\ref{sdeq}).

\subsection{Analytic structure}
\label{analsection}
While we are only interested in $\langle \phi^{2k}\rangle$  and their various approximations for positive real values of the coupling $\lambda$, some of the properties we will uncover can most easily be understood by paying attention to the properties of $\langle \phi^{2k}\rangle$ on the whole complex $\lambda$ plane. 

From (\ref{exact2k}) and the analytic properties of Bessel functions $K_\nu(z)$ on the complex plane \cite{watson}, it follows that all $\langle \phi^{2k} \rangle$ have a branch cut, that is conventionally placed along the negative real axis. Since $K_{\frac{1}{4}}(\frac{1}{8\lambda})$ appears in the denominator of (\ref{exact2k}), zeros of $K_{\frac{1}{4}}(\frac{1}{8\lambda})$ would imply poles of $\langle \phi^{2k} \rangle$ in the complex plane. It is well known that for any real $\nu$, $K_\nu(z)$ doesn't have poles with Re $(z)\geq 0$ \cite{watson}. On the semiplane with Re $z<0$, if $\nu-\frac{1}{2}$ is not an integer $K_\nu(z)$ has $2m$ zeros, where $2m$ is the even integer closest to $\nu-\frac{1}{2}$ \cite{watson}. In the case at hand, the closest integer to $\frac{1}{4}-\frac{1}{2}=-\frac{1}{4}$ is 0, so $K_{\frac{1}{4}}(z)$ has no zeros, and $\langle \phi^{2k} \rangle$ have no poles. Thus, all 2k-point functions $\langle \phi^{2k} \rangle$ are holomorphic away from the branch cut along the negative real axis.

\begin{figure}[h]
\centering
\includegraphics[scale=0.4]{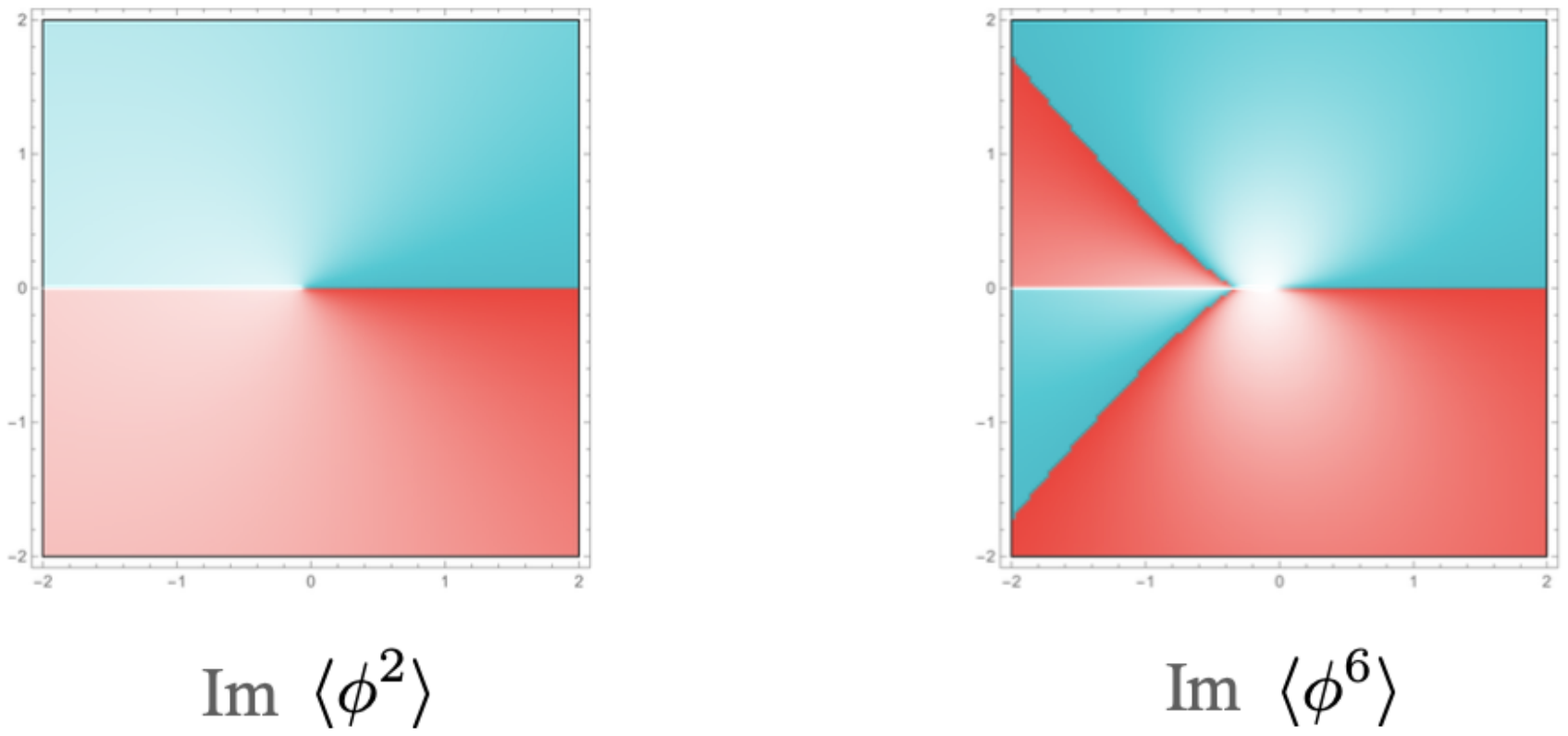}
\caption{Imaginary parts of $\langle \phi^2 \rangle $ and $\langle \phi^6 \rangle$ in the complex $\lambda$ plane. The sign of  $\mathfrak{Im}\, \langle \phi^2\rangle $ is minus the sign of $\mathfrak{Im} \, \lambda$.  The plot for $\mathfrak{Im} \langle \phi^4 \rangle $ is qualitatively similar. On the other hand, the sign of  $\mathfrak{Im}\, \langle \phi^6\rangle $ changes across the upper half plane. The same is true for  $\langle \phi^{2k}\rangle $, $k\geq 3$. This implies that $\langle \phi^6\rangle $ and higher n-point functions are not Stieltjes functions.}
\label{herglotz}
\end{figure}

For the purposes of this paper, the next relevant question about $\langle \phi^{2k} \rangle (\lambda)$ is which of these functions are Stieltjes functions. We review the definition and basic properties of Stieltjes functions in appendix \ref{stieltjesapp}. Here it suffices to recall that a Stieltjes function can be written as
\be
S(z)= \int_0^\infty dt \frac{\rho(t)}{1+zt}
\ee
with $\rho(t)$ a positive function. As reviewed in appendix \ref{stieltjesapp}, a way to prove that a function $S(z)$ is Stieltjes is to prove that it satisfies the four properties of (\ref{stieltjescond}). Properties $i)$ to $iii)$ are easily seen to be satisfied by all $\langle \phi^{2k} \rangle$. Property $iv)$ is more delicate: it demands that $-S(z)$ is a Herglotz function, namely sign $(\mathfrak{Im} \, \langle \phi^{2k} \rangle )=- \text{sign } (\mathfrak{Im} \, \lambda )$. While finding an exact expression for $\mathfrak{Im} \,\, \langle \phi^{2k} \rangle $ is rather challenging, from (\ref{strong}) it immediately follows that for large absolute value of $\lambda$
\be
\text{Arg } \langle \phi^{2k} \rangle \rightarrow \,\, - \frac{k}{2} \text{ Arg  } \lambda
\label{argrelation}
\ee
For $k>2$, this implies that as Arg $\lambda$ increases from 0 to $\pi$, the argument of  $\langle \phi^{2k} \rangle$ changes sign before Arg $\lambda$ reaches $\pi$. This demonstrates that  $\langle \phi^{2k} \rangle$ with $k>2$ are not (minus) Herglotz functions, and therefore they are not Stieltjes functions. The difference of behavior  is illustrated in figure (\ref{herglotz}). 

We will now demonstrate that $\langle \phi^{2}\rangle$ and $\langle \phi^{4}\rangle$ are Stieltjes functions.  Since $\overline{\langle \phi^{2k}\rangle (\lambda)}=\langle \phi^{2k}\rangle (\bar \lambda)$, it is enough to consider the behavior on the upper half plane. Since $\langle \phi^{2k} \rangle$ are holomorphic away from the branch cut, their imaginary parts $\mathfrak{Im} \,\langle \phi^{2k} \rangle$ are harmonic functions away from the branch cut. We will argue that  $\mathfrak{Im} \, \langle \phi^{2}\rangle$ and $\mathfrak{Im} \, \langle \phi^{4}\rangle$ are negative definite on a contour defined by a line just above the real axis and the big semicircle at infinity on the upper half-plane. Then, since harmonic functions on a domain have their maximum at the boundary of the domain,  $\mathfrak{Im} \, \langle \phi^{2}\rangle$ and $\mathfrak{Im} \, \langle \phi^{4}\rangle$ are negative definite in the upper half-plane, concluding the proof that $\langle \phi^{2}\rangle$ and $\langle \phi^{4}\rangle$ are Stieltjes functions.

Let's start with $\langle \phi^{2}\rangle$. From (\ref{argrelation}) we deduce that $\mathfrak{Im} \, \langle \phi^{2}\rangle$ is negative on the big semicircle on the upper half-plane. On the positive real axis, $\langle \phi^{2}\rangle$ is real, so we compute its imaginary part just above the real axis,
\be
\langle \phi^{2}\rangle(|\lambda|+i 0^+) = \langle \phi^{2}\rangle(|\lambda|)-i \frac{xK_{\frac{3}{4}}(x)K_{\frac{5}{4}}(x)-(1+x)K_{\frac{1}{4}}(x)^2}{K_{\frac{1}{4}}(x)^2} \frac{\mathfrak{Im} \,\, \lambda}{2 |\lambda|^2} +\dots
\ee
where $x=\mathfrak{Re}\,\, \frac{1}{8\lambda}$. Now, for $|\nu|<\frac{1}{2}$ and $x>0$, the improved Tur\'an inequality
\be
(1+x)K_\nu^2(x) < x K_{1-\nu}(x) K_{1+\nu}(x)
\ee
that appears in eq. (3.5) of \cite{baricz} proves that $\mathfrak{Im}\,\, \langle \phi^{2}\rangle (|\lambda|+i0^+)$ is negative just above the positive real axis. On the negative real axis, $\langle \phi^2 \rangle$ has a branch cut, and its imaginary part presents a discontinuity. A straightforward computation yields
\be
\mathfrak{Im} \,\,  \langle \phi^{2}\rangle (-|x|+i 0^+) = -\frac{8}{\Gamma\left(\frac{1}{4}\right)\Gamma\left(\frac{3}{4}\right)} \frac{1}{\text{I}_{\frac{1}{4}}(x)^2+\text{I}_{-\frac{1}{4}}(x)^2} 
\label{discphi2}
\ee
where $I_\nu(z)$ are modified Bessel functions of the first kind. (\ref{discphi2}) is manifestly negative, so $\mathfrak{Im} \,\, \langle \phi^{2}\rangle$ is negative on the contour and thus in the interior. This concludes the proof that $\langle \phi^{2}\rangle$ is a Stieltjes function. Having established that $\langle \phi^2 \rangle$ is a Stieltjes function, it follows from the discontinuity (\ref{discphi2}) that $\langle \phi^2 \rangle $ can be written in Stieltjes form,
\be
\langle \phi^2 \rangle (\lambda)=\int_0^\infty dt \, 
\frac{4\sqrt{2}}{\pi^2 t}
\frac{1}{I_{\frac{1}{4}}\left(\frac{t}{8}\right )^2+I_{-\frac{1}{4}}\left(\frac{t}{8}\right )^2} \frac{1}{1+t\lambda}
\label{spectral2}
\ee
As a check, if one expands $\frac{1}{1+t\lambda}$ as a geometric series and performs the integrals for the moments
\be
\int_0^\infty dt \,\, t^n \,\,
\frac{4\sqrt{2} }{\pi^2 t}
\frac{1}{I_{\frac{1}{4}}\left(\frac{t}{8}\right )^2+I_{-\frac{1}{4}}\left(\frac{t}{8}\right )^2}
\ee
one should recover the coefficients of the asymptotic expansion $\langle \phi^2 \rangle_{\text{weak}}$, eq. (\ref{weak}),
\be
\langle \phi^2 \rangle_{\text{weak}} = 1-3\lambda+24 \lambda^2-297 \lambda^3 +4896 \lambda^4 - \dots
\label{twoweak}
\ee
and we have numerically checked that this is the case for the first values of $n$. We can now repeat the argument for $\langle \phi^4 \rangle$,
\be
\langle \phi^4 \rangle=\frac{1}{2\lambda^2}  \frac{(1+2\lambda)K_{\frac{1}{4}} (\frac{1}{8\lambda})-K_{\frac{3}{4}} (\frac{1}{8\lambda})}{K_{\frac{1}{4}} (\frac{1}{8\lambda})}
\ee
Again, (\ref{argrelation}) shows that $\langle \phi^4 \rangle$ has negative imaginary part on the big semicircle in the upper half-plane. Just above the positive real axis we find,
\be
\langle \phi^4 \rangle(|\lambda|+i 0^+)= \langle \phi^4 \rangle(|\lambda|)-i \frac{(x^2+2x+\frac{1}{4})K_{\frac{1}{4}}(x)^2-\frac{3x}{2} K_{\frac{3}{4}}(x) K_{\frac{1}{4}}(x)-x^2 K_{\frac{3}{4}}(x)^2}{K_{\frac{1}{4}}(x)^2} \frac{4 \mathfrak{Im} \,\, \lambda}{|\lambda|^2}+\dots
\label{phi4neg}
\ee
where again $x=\mathfrak{Re} \,\, \frac{1}{8\lambda}$. The proof that the numerator in (\ref{phi4neg}) is positive definite for $x>0$ is slightly more involved that in the previous case; it is due to Javier Segura, and it appears in appendix \ref{appsegura}. On the negative real axis we find
\be
\mathfrak{Im} \,\,  \langle \phi^{4}\rangle (-|\lambda|+i 0^+) = -\frac{8}{|\lambda| \Gamma\left(\frac{1}{4}\right)\Gamma\left(\frac{3}{4}\right)} \frac{1}{\text{I}_{\frac{1}{4}}(x)^2+\text{I}_{-\frac{1}{4}}(x)^2} 
\label{discphi4}
\ee
which is manifestly negative, concluding the argument that $\langle \phi^{4}\rangle$ is a Stieltjes function. Indeed, from the expression above we arrive at
\be
\langle \phi^4 \rangle (\lambda)=\int_0^\infty dt \, 
\frac{4\sqrt{2}}{\pi^2}
\frac{1}{I_{\frac{1}{4}}\left(\frac{t}{8}\right )^2+I_{-\frac{1}{4}}\left(\frac{t}{8}\right )^2} \frac{1}{1+t\lambda}
\label{spectral4}
\ee
The spectral densities of $\langle \phi^2\rangle$ and $\langle \phi^4 \rangle$ in (\ref{spectral2}) and (\ref{spectral4}) differ just by a power of $t$. This had to be the case, since the SD equation $\lambda \langle \phi^4\rangle =-\langle\phi^2 \rangle +1$ together with (\ref{twoweak}) imply that
\be
\langle \phi^4 \rangle_{\text{weak}} = 3-24 \lambda+297 \lambda^2-4896 \lambda^3 +\dots
\ee
so the coefficients in $\langle  \phi^2 \rangle_{\text{weak}}$ and $\langle \phi^4 \rangle_{\text{weak}}$, coincide up to a shift in the powers of $\lambda$.
 
We have proved that $\langle \phi^2 \rangle$ and $\langle \phi^4 \rangle$ are Stieltjes functions, and higher n-point functions are not, but there is an interesting twist: we are now going to show that higher n-point functions are products of a finite number of Stieltjes functions. In section (8.6) of \cite{Bender:1999box}, it is pointed out that if $f_n(\lambda)$ is a Stieltjes function, then
\be
f_{n+1}(\lambda)=\frac{f_n(0)-f_n(\lambda)}{\lambda f_n(\lambda)}
\ee
is also a Stieltjes function, so given a Stieltjes function, one obtains by iteration an infinite family of Stieltjes functions. In the case at hand, taking $f_0(\lambda)=\langle \phi^2 \rangle$ and using the SD equations, it can be proven by induction that $f_k(\lambda)=\frac{\langle \phi^{2k+2}\rangle}{\langle \phi^{2k}\rangle}$ are Stieltjes functions for $k\geq 1$. An indication that this is a plausible result is that from (\ref{strong}) we learn that
\be
\text{Arg } \frac{\langle \phi^{2k+2}\rangle}{\langle \phi^{2k}\rangle} \rightarrow -\frac{1}{2} \text{Arg } \lambda
\ee
as $|\lambda |\to \infty$, so the imaginary part doesn't change sign on the upper or lower half planes, at least for large enough $|\lambda|$. This result implies that, while $\langle \phi^6 \rangle$ and higher n-point functions are not Stieltjes functions, they are products of a finite number of Stieltjes functions, {\it e.g.}
\be
\langle \phi^6 \rangle = \langle \phi^2 \rangle \,\, \frac{\langle \phi^{4}\rangle}{\langle \phi^{2}\rangle}
\,\, \frac{\langle \phi^{6}\rangle}{\langle \phi^{4}\rangle}.
\ee
Note that in this regard $\langle \phi^4 \rangle$ plays a peculiar role, since it is a Stieltjes function, and can also be written as a product of Stieltjes functions,
\be
\langle \phi^4 \rangle = \langle \phi^2 \rangle \,\, \frac{\langle \phi^{4}\rangle}{\langle \phi^{2}\rangle}.
\ee
The fact that $\langle \phi^6 \rangle$ and higher n-point functions are not Stieltjes functions but can be written as products of Stieltjes functions will provide a useful perspective to understand their rational approximants, to which we now turn our attention.

\subsection{Schwinger-Dyson approximants}
For the theory at hand, we have just reviewed that one can obtain the exact n-point functions, and check that they satisfy the SD equations. This will hardly ever be the case for more complicated QFTs. Bearing this motivation in mind, we now turn the question around, and ask whether we can use the SD equations to find the n-point functions $\langle \phi^{2k}\rangle $. The linearity of the Schwinger-Dyson equations (\ref{sdeq}) turns out to be key in this endeavour, as it allows to write them in matrix form,
\be
\begin{pmatrix}
1 & 0 & 0 & 0 & \dots \\
-1 & 1 & \lambda & 0 & \dots \\
0 & -3 & 1 & \lambda & \dots \\
0 & 0 &-5 & 1 & \dots \\
\vdots & \vdots & \vdots & \vdots & \ddots
\end{pmatrix}
\begin{pmatrix}
\langle 1 \rangle  \\
\langle \phi^2 \rangle \\
\langle \phi^4 \rangle \\
\langle \phi^6 \rangle \\
\vdots 
\end{pmatrix}
=
\begin{pmatrix}
1 \\
0 \\
0 \\
0 \\
\vdots 
\end{pmatrix}
\label{sdmatrix}
\ee
This matrix form of the SD equations appeared already in \cite{Bender:1988bp}. Very recently, this equation has been independently rederived in \cite{Konosu:2024zrq} from a superficially different approach, the $A_\infty$ homotopy algebra formulation of QFT\footnote{The matrix in \cite{Konosu:2024zrq} is larger, as it acts on a vector of all n-point functions, not just the even ones. Since the odd n-point functions vanish anyway, one can discard the relevant rows and columns, and the resulting matrix is precisely the one in  (\ref{sdmatrix}).}.
Applying Cramer's rule to (\ref{sdmatrix}) we arrive at
\be
\langle \phi^{2k} \rangle_{\text{SD}} = (2k-1)!!
\frac{
\begin{vmatrix}
1 & \lambda  & 0 & 0 & \dots \\
-(2k+3) & 1 & \lambda & 0 & \dots \\
0 & -(2k+5) & 1 &  \lambda & \dots \\
0 & 0 & -(2k+7) & 1 & \dots \\
\vdots & \vdots & \vdots & \vdots  & \ddots 
\end{vmatrix}}{\begin{vmatrix}
1 & \lambda  & 0 & 0 & \dots \\
-3 & 1 & \lambda & 0 & \dots \\
0 & -5 & 1 &  \lambda & \dots \\
0 & 0 & -7 & 1 & \dots \\
\vdots & \vdots & \vdots & \vdots  & \ddots 
\end{vmatrix}}
\label{cramer}
\ee
By truncating the initial semi-infinite matrix in (\ref{sdmatrix}) to a finite square $n\times n$ matrix, the resulting determinants in (\ref{cramer}) yield finite degree polynomials in $\lambda$, so the truncated SD equations are solved by rational functions, that we refer to as the Schwinger-Dyson rational approximants $\langle \phi^{2k} \rangle^n_{\text{SD}}$. The matrices in the numerator and denominator of (\ref{cramer}) are both tridiagonal, so their determinants are given by three-term recurrence relations. We define families of polynomials $p_n^{(k)}(\lambda)$ for $k=0,1,2,\dots$ and $n=0,1,2,\dots$ by the following recursion relations,

\be
p_n^{(k)}(\lambda)=
\begin{cases}
0 & n<k+1 \\
1 & n=k+1 \\
p_{n-1}^{(k)}(\lambda)+(2n-3) \lambda p_{n-2}^{(k)} & n > k+1
\end{cases}
\label{recurp}
\ee
Since the denominators $p_n^{(0)}(\lambda)$ will have a prominent role in the subsequent discussion, we introduce the simpler notation $q_n(\lambda)=p_n^{(0)}(\lambda)$. Then
\be
\highlight{\langle \phi^{2k} \rangle^n_{\text{SD}}= (2k-1)!! \,\, \frac{p_n^{(k)}(\lambda)}{q_{n}(\lambda)}}
\label{thehaapprox}
\ee

This is the main result of this subsection. In table (\ref{tableofsdapp}), we illustrate this result with the first few SD approximants.

\begin{table}[h!]
\begin{center}
\begin{tabular}{ c | c c c c}
 & $\langle \phi^2\rangle^n_{\text{SD}}$ & $\langle \phi^4\rangle^n_{\text{SD}}$ & $\langle \phi^6\rangle^n_{\text{SD}}$ & $\langle \phi^8\rangle^n_{\text{SD}}$ \\
 \hline
n=2 & 1 & 0 & 0 & 0 \\
  &&&& \\
n=3 & $\frac{1}{1+3\lambda}$ & $ \frac{3}{1+3\lambda}$ & 0 & 0 \\
  &&&& \\
n=4 & $\frac{1+5\lambda}{1+8\lambda}$ & $\frac{3}{1+8\lambda}$ & $\frac{15}{1+8\lambda}$ & 0 \\
  &&&& \\
n=5 & $\frac{1+12\lambda}{1+15\lambda+21\lambda^2}$ & $\frac{3+21\lambda}{1+15\lambda+21\lambda^2}$ & $\frac{15}{1+15\lambda+21\lambda^2}$ & $\frac{105}{1+15\lambda+21\lambda^2}$ 
\end{tabular}
\end{center}
\caption{The first Schwinger-Dyson approximants for various $\langle \phi^{2k} \rangle$.}
\label{tableofsdapp}
\end{table}

If we denote by $[L,M]$ a rational approximant given by a numerator polynomial of degree L divided by a denominator polynomial of degree M, note that the $\langle \phi^{2} \rangle^n_{\text{SD}}$ alternate in being [N,N] and [N,N+1] approximants. As for $\langle \phi^{4} \rangle^n_{\text{SD}}$, the non-zero approximants are always of the $[N,N+1]$ type, with two different approximants per N. In general,  $\langle \phi^{4k} \rangle^n_{\text{SD}}$ are [N,N+k] approximants, and $\langle \phi^{4k-2} \rangle^n_{\text{SD}}$ alternate in being [N,N+k-1] and [N,N+k] approximants.

Some comments are in order:
\begin{enumerate}
\item{By construction, the rational approximants (\ref{thehaapprox}) satisfy the first $n-1$ Schwinger-Dyson equations.}
\item{For a given order of the truncation of the matrix in (\ref{sdmatrix}), the approximants to all $\langle \phi^{2k} \rangle$ share the same denominator, $q_n(\lambda)$.}
\item{All the coefficients of the $p_n^{(k)}(\lambda)$ and $q_n(\lambda)$ polynomials are manifestly positive. Therefore, the rational approximants $\langle \phi^{2k} \rangle^n_{\text{SD}}$ have neither zeros nor poles for real positive values of the coupling $\lambda$.}
\item{Since for $\lambda>0$, $q_n(\lambda)> p_n^{(k)}(\lambda)$, for positive values of the coupling these rational approximants are bounded by
$$
0< \langle \phi^{2k}\rangle^n_{\text{SD}} < (2k-1)!!, \hspace{1cm} \text{for } \lambda >0.
$$
}

\end{enumerate}

One can Taylor expand these finite degree rational approximants $\langle \phi^{2k} \rangle^n_{\text{SD}}$ to obtain an infinite power series in $\frac{\lambda}{m^4}$. As the size of the truncation increases, the number or terms in these infinite power series that agree with $\langle \phi^{2k} \rangle_{\text{weak}}$ also increases. One might then be tempted to leap to the conclusion that $\langle \phi^{2k} \rangle^n_{\text{SD}}$ are no better at capturing the strong coupling regime of $\langle \phi^{2k} \rangle$  than the asymptotic series $\langle \phi^{2k} \rangle_{\text{weak}}$. If we are reading \cite{Bender:1988bp} correctly, this appears to be the claim of \cite{Bender:1988bp} with regards to $\langle \phi^{2k} \rangle^n_{\text{SD}}$. On the other hand, the authors of \cite{Konosu:2024zrq} claim, but do not prove, that as one takes the size of the truncation to infinity, $\langle \phi^{2k} \rangle^n_{\text{SD}}$ tend to the exact answer, even in the strong coupling regime. In section \ref{convergence}, we will prove the convergence of $\langle \phi^{2k} \rangle^n_{\text{SD}}$
to $\langle \phi^{2k} \rangle$ for all positive values of the coupling.

\section{Comparison with Pad\'e approximants}
\label{comparison}
The appearance of rational approximants $\langle \phi^{2k} \rangle^n_{\text{SD}}$ as solutions of the truncated SD equations immediately raises the question of whether these rational functions are Pad\'e approximants to some power series. Since - up to an overall $1/m^{2k}$ factor  - they are ratios of polynomials in $\frac{\lambda}{m^4}$, the first thought is that they might be Pad\'e approximants for $\langle \phi^{2k} \rangle_{\text{weak}}$, eq. (\ref{weak}). An immediate difference is that usually the Pad\'e approximants are derived from a perturbative series, $\langle \phi^{2k} \rangle_{\text{weak}}$ in our case. On the other hand, the derivation of the rational approximants (\ref{thehaapprox}) from the truncated Schwinger-Dyson equations completely bypasses the need for a perturbative series.

In this section we compare the new rational approximants (\ref{thehaapprox}) with the relevant  Pad\'e approximants to $\langle \phi^{2k} \rangle_{\text{weak}}$. The outcome of the comparison is different for different 2k-point functions:  $\langle \phi^{2} \rangle^n_{\text{SD}}$ are indeed certain Pad\'e approximants for $\langle \phi^{2} \rangle_{\text{weak}}$. For the 4-point function, $\langle \phi^{4} \rangle^n_{\text{SD}}$ are Pad\'e approximants to $\langle \phi^{4} \rangle_{\text{weak}}$ for odd n, but not for even n. Finally, for $k>2$, $\langle \phi^{2k} \rangle^n_{\text{SD}}$ are not Pad\'e approximants to $\langle \phi^{2k} \rangle_{\text{weak}}$; instead they are the product of Pad\'e approximants to series of the form $\frac{\langle \phi^{2m+2}\rangle_{\text{weak}}}{\langle \phi^{2m}\rangle_{\text{weak}}}$.

Let's start by recalling the definition of Pad\'e approximants \cite{baker, Bender:1999box}. For any power series 
\be
f(z)=\sum_{i=0}^\infty c_i z^i
\label{powerseries}
\ee
its $[L,M]$ Pad\'e approximant is given by the quotient of a polynomial of degree L divided by a polynomial of degree M,
\be
f(z)^{[L,M]}=\frac{a_0+a_1z+\dots +a_L z^L}{b_0+b_1 z+\dots +b_M z^M}
\ee
such that 
\be
f(z)-f(z)^{[L,M]}={\cal O}(z^{L+M+1})
\ee
so they are the best rational approximation to the power series. From our perspective, the main reason to be interested in Pad\'e approximants is that there are instances where the convergence properties of a family of Pad\'e approximants are better than those of the original power series \cite{baker,  Bender:1999box}. 

Since $\langle \phi^{2k} \rangle^n_{\text{SD}}$ are - up to an overall power of $1/m^2$ - functions of $\frac{\lambda}{m^4}$, it makes sense to compare them with the Pad\'e approximants of $\langle \phi^{2k} \rangle_{\text{weak}}$. As $\langle \phi^{2k} \rangle^n_{\text{SD}}$ are rational approximants with specific degrees for the numerator and denominator, we compare them with Pad\'e approximants of the same degrees. The results appear in table (\ref{SDpade}).  For a given $\langle \phi^{2k}\rangle$, in the cases where the two approximants disagree, one can check that as the degrees of the approximants increase, the number of coincident terms in the power expansions of the two approximants also increases.
 
\begin{table}[h!]
\begin{center}
\begin{tabular}{c || c  c  c  c c c}
$\langle \phi^2\rangle_{\text{SD}}$ & 1 & $\frac{1}{1+3\lambda}$ & $\frac{1+5\lambda}{1+8\lambda}$ &  $\frac{1+12\lambda}{1+15\lambda+21\lambda^2}$ & $\frac{1+21\lambda+45 \lambda^2}{1+24\lambda+93 \lambda^2}$ & $\dots$ \\
 &&&&&&\\
$\langle \phi^2\rangle_{\text{Pad\'e}}$ & 1 & $\frac{1}{1+3\lambda}$ & $\frac{1+5\lambda}{1+8\lambda}$ &  $\frac{1+12\lambda}{1+15\lambda+21\lambda^2}$ & $\frac{1+21\lambda+45 \lambda^2}{1+24\lambda+93 \lambda^2}$ & \dots \\
 &&&&&&\\
\hline
$\langle \phi^4\rangle_{\text{SD}}$ & $3\frac{1}{1+3\lambda}$ & $3\frac{1}{1+8\lambda}$ & 3 $\frac{1+7\lambda}{1+15\lambda+21\lambda^2}$ & $3\frac{1+16\lambda}{1+24\lambda+93\lambda^2}$ & $3\frac{1+27\lambda+77\lambda^2}{1+35\lambda+258\lambda^2+231\lambda^3}$ & \dots \\
 &&&&&&\\
$\langle \phi^4\rangle_{\text{Pad\'e}}$  & & $3\frac{1}{1+8\lambda}$ & &$3\frac{1+16\lambda}{1+24\lambda+93\lambda^2}$  &
  & \dots\\
 &&&&&&\\
\hline
$\langle \phi^6\rangle_{\text{SD}}$ & $\frac{15}{1+8\lambda}$ & $\frac{15}{1+15\lambda+21\lambda^2}$ & $\frac{15+135\lambda}{1+24\lambda+93\lambda^2}$& $\frac{15+300\lambda}{1+35\lambda+258\lambda^2+231\lambda^3}$& $\frac{15+495\lambda+1755\lambda^2}{1+48\lambda+570\lambda^2+1440\lambda^3}$&  \dots \\

&  &   &  &&&\\

$\langle \phi^6\rangle_{\text{Pad\'e}}$ & $\frac{15}{1+15\lambda}$ & $\frac{15}{1+15\lambda-42 \lambda^2}$ &  $\frac{15+\frac{765}{2} \lambda}{1+\frac{81}{2}\lambda+\frac{681}{2}\lambda^2}$ &  $\frac{15+\frac{7245}{17}\lambda}{1+\frac{738}{17}\lambda+\frac{6531}{17}\lambda^2-\frac{2079}{17}\lambda^3}$ & $\frac{15+1145 \lambda + 18330 \lambda^2}{1+\frac{274}{3}\lambda+2325 \lambda^2 +16195 \lambda^3}$& \dots
\end{tabular}
\caption{Table comparing the first non-zero Schwinger-Dyson approximants to $\langle \phi^2 \rangle$, $\langle \phi^4 \rangle$, $\langle \phi^6 \rangle$ and the relevant Pad\'e approximants to $\langle \phi^{2,4,6} \rangle_{\text{weak}}$.} 
\label{SDpade}
\end{center}
\end{table}

\subsection{$\langle \phi^2 \rangle_{\text{SD}} = \langle \phi^2 \rangle_{\text{Pad\'e}}$}
\label{twotwo}
For $\langle \phi^2 \rangle$, the comparison of the first approximants in table \ref{SDpade} suggests that its Schwinger-Dyson approximants coincide with the [N,N] and [N,N+1] Pad\'e approximants to $\langle \phi^2 \rangle_{\text{weak}}$. We will now prove rigorously that this is the case. The proof we are about to present takes advantage of the knowledge of the exact n-point functions (\ref{exactparabolic}); knowledge of the exact n-point function is rather uncommon, so in appendix \ref{appa} we present a second proof that makes use only of the perturbative series $\langle \phi^2 \rangle_{\text{weak}}$ (\ref{weak}).

An efficient way to obtain the $[N,N]$ and $[N,N+1]$ Pad\'e approximant of a series is to write the series as a continued fraction \cite{baker, Bender:1999box}. For this reason, we will start by producing a continued fraction presentation for $\langle \phi^2\rangle_{\text{weak}}$. In fact, it is equally simple and useful to present such continued fraction for all Stieltjes series $\frac{\langle \phi^{2k+2} \rangle_{\text{weak}}}{\langle \phi^{2k} \rangle_{\text{weak}}}$.

Recall that the exact n-point functions can be written in terms of parabolic cylinder functions $U_a(x)$, eq. (\ref{exactparabolic}). Then, from the continued fraction for $U_a(x)/U_{a-1}(x)$, eq. (16.5.7) in \cite{cuyt} we immediately arrive at \cite{Garcia:1996np}
\be
\frac{\langle \phi^{2k+2}\rangle_{\text{weak}}}{\langle \phi^{2k}\rangle_{\text{weak}}}=
(2k+1)
\cfrac{1}{1+\cfrac{(2k+3)\lambda}{1+\cfrac{(2k+5)\lambda}{1+\cfrac{(2k+7)\lambda}{1+\dots}}}}
\ee
From this continued fraction, and applying theorem 4.2.1 of \cite{baker} we read off the [N,N] and [N,N+1] Pad\'e approximants to $\frac{\langle \phi^{2k+2}\rangle_{\text{weak}}}{\langle \phi^{2k}\rangle_{\text{weak}}}$; they are given by $(2k+1)\frac{A_n}{B_n}$ with
\be
\begin{split}
A_0=0, \hspace{.5cm} A_1=1, \hspace{.5cm} & A_n=A_{n-1}+(2k+2n-1)\lambda A_{n-2}, \,\,\, n\geq 2 \\
B_0=1, \hspace{.5cm} B_1=1,  \hspace{.5cm} & B_n=B_{n-1}+(2k+2n-1)\lambda B_{n-2}, \,\,\, n\geq 2 \\
\end{split}
\ee
Comparing with the definition of the $p_n^{(k)}(\lambda)$ polynomials (\ref{recurp}), we deduce that the [N,N] and $[N,N+1]$ Pad\'e approximants for  $\frac{\langle \phi^{2k+2}\rangle_{\text{weak}}}{\langle \phi^{2k}\rangle_{\text{weak}}}$ are given by
\be
(2k+1)\frac{p_n^{(k+1)}(\lambda)}{p_n^{(k)}(\lambda)}
\label{padeforquot}
\ee
In particular, setting k=0, this proves that the $\langle \phi^2 \rangle^n_{\text{SD}}$ approximants are equal to the [N,N] and [N,N+1] Pad\'e approximants to $\langle \phi^2 \rangle_{\text{weak}}$.

\subsection{SD approximants as products of Pad\'e approximants}
Recall that at the end of section (\ref{analsection}) we argued that while $\langle \phi^{2k}\rangle$ are not Stieltjes functions for $k>2$, they are products of Stieltjes functions,
\be
\langle \phi^{2k} \rangle =
\frac{\langle \phi^{2k} \rangle}{\langle \phi^{2k-2} \rangle} \frac{\langle \phi^{2k-2} \rangle}{\langle \phi^{2k-4} \rangle} 
\dots
\frac{\langle \phi^{4} \rangle}{\langle \phi^{2} \rangle} \langle \phi^{2} \rangle.
\label{prodstielagain}
\ee
Our result (\ref{padeforquot}) for the [N,N] and [N,N+1] Pad\'e approximants of the Stieltjes functions $\frac{\langle \phi^{2k+2}\rangle}{\langle \phi^{2k}\rangle}$ provides a complementary perspective on the Schwinger-Dyson approximants introduced in this work. Namely, since $\langle \phi^6 \rangle$ and higher n-point functions are not Stieltjes functions, many results for Pad\'e approximants to Stieltjes functions \cite{baker, Bender:1999box} don't apply to them. Moreover, as seen in table (\ref{SDpade}) and its extension to higher orders, the Pad\'e approximants for these higher n-point functions have spurious zeros and poles on the positive real axis, that don't correspond to actual zeros or poles of the exact functions. On the other hand, since we can think of these functions as products of Stieltjes functions, eq. (\ref{prodstielagain}), we can construct rational approximants given by products of Pad\'e approximants to each of the Stieltjes functions, eq. (\ref{padeforquot}), and these products are no other than the Schwinger-Dyson approximants 
\be
\langle \phi^{2k} \rangle_{\text{SD}}^n = (2k-1)!! \frac{p_n^{(k)}(\lambda)}{p_n^{(0)}(\lambda)}=
(2k-1) \frac{p_n^{(k)}(\lambda)}{p_n^{(k-1)}(\lambda)} \, (2k-3) \frac{p_n^{(k-1)}(\lambda)}{p_n^{(k-2)}(\lambda)}\,\,
\dots 1 \frac{p_n^{(1)}(\lambda)}{p_n^{(0)}(\lambda)}
\label{productofpade}
\ee
This point of view provides an alternative explanation for the good convergence properties of the Schwinger-Dyson approximants, which we discuss next.


\section{Convergence of the Schwinger-Dyson approximants}
\label{convergence}
In this section we are going to prove that for a given 2k-point function $\langle \phi^{2k} \rangle$, as we take the size of the truncation $n$ to infinity, the family of approximants $\langle \phi^{2k} \rangle^n_{\text{SD}}$ converge to $\langle \phi^{2k} \rangle$, for positive coupling $\lambda$. The common key ingredient of the two proofs that we will present is the demonstation that the Pad\'e approximants (\ref{padeforquot}) to $\frac{\langle \phi^{2k+2}\rangle_{\text{weak}}}{\langle \phi^{2k} \rangle_{\text{weak}}}$ converge to $\frac{\langle \phi^{2k+2}\rangle}{\langle \phi^{2k} \rangle}$. This is equivalent to showing that the Stieltjes function $\frac{\langle \phi^{2k+2}\rangle}{\langle \phi^{2k} \rangle}$ is the unique Stieltjes function that has $\frac{\langle \phi^{2k+2}\rangle_{\text{weak}}}{\langle \phi^{2k} \rangle_{\text{weak}}}$ as Stieltjes series. 

In appendix \ref{appconve} we will present a third proof of convergence, using the fact that all  $\langle \phi^{2k} \rangle^n_{\text{SD}}$ share denominators $q_n(\lambda)$ with $\langle \phi^{2} \rangle^n_{\text{SD}}$. This fact and the similarity of the recursive relations will allow us to adapt familiar arguments for convergence of Pad\'e approximants of Stieltjes series \cite{baker, Bender:1999box} to the novel approximants introduced in this work, and prove their convergence.

Before we delve into the arguments, let's present a simple but far-reaching observation, that will help develop an intuition on how these approximants converge to the limit function, and what's the most delicate point in the proofs. For fixed $\langle \phi^{2k} \rangle$, each approximant is between the previous two. Indeed, from
\be
\frac{p_n^{(k)}(\lambda)}{q_{n}(\lambda)} = \frac{p_{n-1}^{(k)}(\lambda) + (2n-3) \lambda \, p_{n-2}^{(k)}(\lambda)}{q_{n-1}(\lambda) + (2n-3) \lambda \, q_{n-2}(\lambda)} 
\ee
we deduce that for $\lambda >0$


\be
\text{min} \left ( \langle \phi^{2k}\rangle^{n-1}_{\text{SD}}, \langle \phi^{2k}\rangle^{n-2}_{\text{SD}} \right)
\, < \,  \langle \phi^{2k}\rangle^{n}_{\text{SD}} \, < \, \text{max} \left ( \langle \phi^{2k}\rangle^{n-1}_{\text{SD}}, \langle \phi^{2k}\rangle^{n-2}_{\text{SD}} \right)
\ee

\begin{figure}[h]
\centering
\includegraphics[scale=0.5]{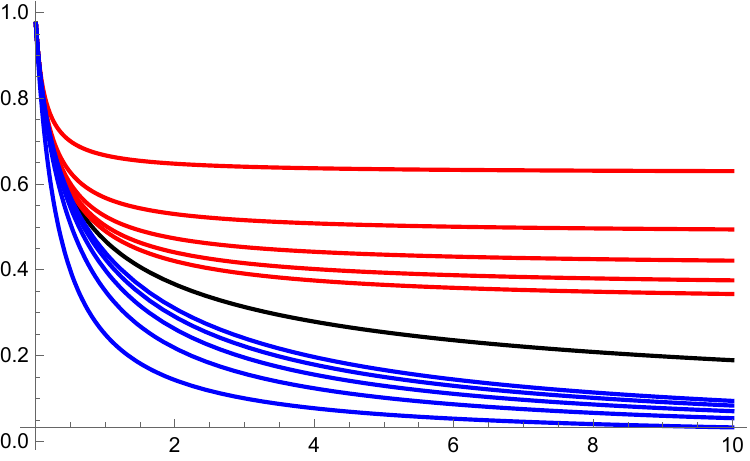}
\caption{Various Schwinger-Dyson approximants to $\langle \phi^2\rangle$. The exact result is depicted in black. The [N,N] approximants - in red - form a decreasing subsequence, while the [N,N+1] approximants - in blue - form an increasing subsequence. Given that these subsequences are bounded and monotonic, it is immediate that each of them has a limit function. What it is less immediate is that both subsequences converge to the same limit function, and that the limit function is the exact n-point function.}
\label{converpade}
\end{figure}

This already implies that for $\lambda>0$, the two subsequences of approximants with even $n$ and with odd $n$ are monotonic and bounded. For odd $k$, the even $n$ subsequence is monotonically decreasing and the  odd $n$ one is monotonically increasing; for even k, the behaviors of the two subsequences are reversed. Since a sequence that is monotonic and bounded is convergent, we learn that for fixed $\lambda>0$, these two subsequences converge to some values, so for $\lambda \in [0,+\infty)$ the subsequences converge to limit functions, see figure (\ref{converpade}). The remaining question is whether these two limit functions are the same or not. It turns out that this question is equivalent to deciding the uniqueness of the Stieltjes moment problem \cite{baker, Bender:1999box, akhiezer, simon}, see appendix \ref{stieltjesapp}. We now want to argue that indeed the two limit functions are the same, so the approximants converge to a single limit function, which moreover is $\langle \phi^{2k}\rangle$. 

We have already proven in section \ref{analsection} that the exact $\frac{\langle \phi^{2k+2}\rangle}{\langle \phi^{2k} \rangle}$ are Stieltjes functions. We have also proven in section \ref{twotwo} that (\ref{padeforquot}) are [N,N] and [N,N+1] Pad\'e approximants for the Stieltjes power series $\frac{\langle \phi^{2k+2}\rangle_{\text{weak}}}{\langle \phi^{2k} \rangle_{\text{weak}}}$. We are then in a position to take advantage of the well developed theory of convergence of Pad\'e approximants to Stieltjes series \cite{baker, Bender:1999box}, see appendix \ref{stieltjesapp} for basic results. The main challenge is to prove that for each fixed $k$, $\frac{\langle \phi^{2k+2}\rangle}{\langle \phi^{2k} \rangle}$ is the unique Stieltjes function that has $\frac{\langle \phi^{2k+2}\rangle_{\text{weak}}}{\langle \phi^{2k} \rangle_{\text{weak}}}$ as asymptotic series. This is by no means an obvious fact; indeed, a given series can be asymptotic to different functions, differing by non-perturbative terms.

To prove uniqueness, we will apply the sufficient criterion for uniqueness given by condition (\ref{uniquecrit})\footnote{Given a generic power series, one has to prove first that it is a Stieltjes power series, before checking for uniqueness. A possible way to do so is via the determinantal conditions, eq. (\ref{determinantal}). For the particular case of $\langle \phi^2 \rangle_{\text{weak}}$, it follows from the results in appendix \ref{appa} that is satisfies the determinantal conditions. However, since we know that these are series asymptotic to Stieltjes functions, we already know that they are Stieltjes series.}. In order to apply this criterion, we need to bound the growth of the coefficients $\mu_n^{(k)}$ in
\be
\frac{\langle \phi^{2k+2}\rangle_{\text{weak}}}{\langle \phi^{2k} \rangle_{\text{weak}}}=
\frac{\sum_{n=0}^\infty \frac{(4n+2k+1)!!}{n!} \left(-\frac{\lambda}{4}\right)^n}{\sum_{n=0}^\infty \frac{(4n+2k-1)!!}{n!} \left(-\frac{\lambda}{4}\right)^n}=\sum_{n=0}^\infty \mu_n^{(k)} (-\lambda)^n
\ee
To do so, denote by $a_n=\frac{(4n+2k+1)!!}{4 ^n n!}$ the coefficients in the numerator, and by $b_n= \frac{(4n+2k-1)!!}{4 ^n n!}$ the ones in the denominator. Then the Cauchy formula for the coefficients of a product of power series reads $a_n=\sum_{i=0}^n \mu_i^{(k)} b_{n-i}$, and from it we find
\be
b_0 \mu_n^{(k)}= a_n-\sum_{i=0}^{n-1} \mu_i^{(k)} b_{n-i}
\ee
We know that $\mu_i^{(k)}$ are the moments of a Stieltjes function, so they are positive - see (\ref{stieltjesmoments}) - and this implies that $\mu_n^{(k)} \leq a_n/b_0$, or
\be
\mu_n^{(k)} \leq \frac{(4n+2k+1)!! }{4^n n! (2k-1)!!}
\ee
which can be very generously bound by
\be
\mu_n^{(k)} \leq \frac{(4n+2k+1)!! }{4^n n! (2k-1)!!} \leq \frac{4}{\pi} \frac{(4+2k)^k}{(2k-1)!!} \,\,e^n \, (2n)!
\ee
This bound is enough to prove that the sufficient criterion for uniqueness, eq. (\ref{uniquecrit}) is satisfied. We then conclude that there is a unique Stieltjes function with  $\frac{\langle \phi^{2k+2}\rangle_{\text{weak}}}{\langle \phi^{2k} \rangle_{\text{weak}}}$ as asymptotic series. Since  $\frac{\langle \phi^{2k+2}\rangle}{\langle \phi^{2k} \rangle}$ is such a function, it is the unique one. Finally, we invoke theorem 5.5.1 in \cite{baker} - see appendix \ref{stieltjesapp} - that implies that the Pad\'e approximants (\ref{padeforquot}) converge to $\frac{\langle \phi^{2k+2}\rangle}{\langle \phi^{2k} \rangle}$ in the cut plane,
\be
\lim_{n \to \infty} (2k+1)\frac{p_n^{(k+1)}(\lambda)}{p_n^{(k)}(\lambda)} = \frac{\langle \phi^{2k+2}\rangle}{\langle \phi^{2k} \rangle} .
\label{quotconve}
\ee
This result provides two quick demonstrations of the convergence of $\langle \phi^{2k} \rangle^n_{\text{SD}}$ to $\langle \phi^{2k} \rangle$. In the first one, we set $k=0$ in (\ref{quotconve}) to conclude that $\langle \phi^{2} \rangle^n_{\text{SD}}$ converges to $\langle \phi^{2} \rangle$. We can then recursively use the Schwinger-Dyson equations, which are satisfied both by the SD approximants and by the exact 2k-point function. For instance,
\be
\lim_{n \to \infty} 
\langle \phi^4\rangle^n_{\text{SD}}  = \lim_{n \to \infty} \frac{\langle \phi^2\rangle^n_{\text{SD}} -1}{\lambda} = \frac{\langle \phi^2\rangle -1}{\lambda} =\langle \phi^4\rangle
\ee
where we have used the SD equations in the first and third step. Then, we can repeat the argument for $\langle \phi^6\rangle^n_{\text{SD}}$ and so on.

A second proof of convergence puts to use that $\langle \phi^{2k} \rangle$ is a product of Stieltjes functions, eq. (\ref{prodstielagain}), and $\langle \phi^{2k} \rangle^n_{\text{SD}}$ is a product of their Pad\'e approximants, eq. (\ref{productofpade}). Then,
\be
\begin{split}
& \lim_{n \to \infty} \langle \phi^{2k} \rangle_{\text{SD}}^n = 
\lim_{n \to \infty}
(2k-1) \frac{p_n^{(k)}(\lambda)}{p_n^{(k-1)}(\lambda)} \, (2k-3) \frac{p_n^{(k-1)}(\lambda)}{p_n^{(k-2)}(\lambda)}\,\,
\dots 1 \frac{p_n^{(1)}(\lambda)}{p_n^{(0)}(\lambda)} \\
& = \frac{\langle \phi^{2k} \rangle}{\langle \phi^{2k-2} \rangle} \frac{\langle \phi^{2k-2} \rangle}{\langle \phi^{2k-4} \rangle} 
\dots
\frac{\langle \phi^{4} \rangle}{\langle \phi^{2} \rangle} \langle \phi^{2} \rangle = \langle \phi^{2k} \rangle.
\end{split}
\ee

\acknowledgments
We would like to thank Keisuke Konosu, Pere Masjuan, Yuji Okawa and Santi Peris for helpful conversations and correspondence. We are particulaly grateful to Javier Segura for providing the proof in appendix \ref{appsegura}. The research of BF is supported in part by grants PID2022-136224NB-C22,  2021-SGR-00872 by AGAUR (Generalitat de Catalunya) and CEX2024-001451-M funded by MICIU/AEI/10.13039/501100011033. 

\appendix

\section{Stieltjes, Stieltjes, Stieltjes} 
\label{stieltjesapp}

In this appendix we collect the definition and basic properties of Stieltjes functions, Stieltjes series and Stieltjes continued fractions. Some references that we have found useful are \cite{baker, Bender:1999box, akhiezer, simon}. 

A Stieltjes function is a function $S(\lambda)$ that can be written as
\be
S(z)=\int_0^\infty \frac{\mu(x) \, dx }{1+x z}
\label{defstieltjes}
\ee
for some $\mu(x)$ that is positive for $x\geq 0$, and such that its moments
\be
\mu_n = \int_0^\infty x^n \mu(x) \, dx
\label{stieltjesmoments}
\ee
are all finite. Note that $\mu_n \geq 0$, and that the integrals in the definition above are over the positive real axis, not the full real axis. Taking the imaginary part of (\ref{defstieltjes}) we learn that a Stieltjes function sastisfies
\be
\text{sign }(\mathfrak{Im} \, S(z) ) = - \text{ sign }(\mathfrak{Im } \, z)  
\ee
so it maps the upper half plane to the lower half plane and viceversa, {\it i.e. } $-S(z)$ must be a Herglotz function. In \cite{Bender:1999box} the following criterion is presented to prove that a function is Stieltjes:

\be
\begin{tabular}{c l}
i) & S(z) is analytic in the cut plane. \\
ii) & S(z)  $\to C$ \text{as} $z\to \infty$ \text{where C is a nonnegative constant.}\\
iii) & \text{S(z) has asymptotic series representation of the form $\sum_{n=0}^\infty a_n(-z)^n$ in the cut plane.} \\
iv) & \text{-S(z) is a Herglotz function.}
\end{tabular}
\label{stieltjescond}
\ee

A power series $\sum_{k=0}^\infty \mu_k(-\lambda)^k$ is a Stieltjes series if $\mu_k$ are the moments of a Stieltjes function. 

Finally, a Stieltjes continued fraction is a continued fraction of the form
\be
\cfrac{a_1}{1+\cfrac{a_2 z}{1+\cfrac{a_3z}{1+\dots}}}
\ee
with all $a_i$ positive. 

By definition, the power series expansion of a Stieltjes function provides a Stieltjes series. Furthermore, according to theorem 5.5.2 in \cite{baker}, a Stieltjes function admits a continued fraction representation of the Stieltjes type. A set of more delicate questions is whether given a power series or a continued fraction there exist a related Stieltjes function, and if so, whether it is unique.

Given a power series $\sum_{k=0}^\infty \mu_k(-\lambda)^k$, deciding whether it is a Stieltjes series is a moment problem \cite{akhiezer, simon}: we need to find out whether there is some positive $\mu(x)$ over the positive real axis such that its moments are $\mu_k$. A necessary and sufficient condition for the  existence of the measure was discussed already by Stieltjes, and it is given by determinantal conditions
\be
D(0,n)=
\begin{vmatrix}
\mu_0 & \mu_1 & \mu_2 &\dots & \mu_n \\
\mu_1 & \mu_2 & \mu_3 & \dots  & \vdots \\
\dots & \dots & \dots & \ddots & \vdots \\
\mu_n & & & & \mu_{2n}
\end{vmatrix} >0, 
\hspace{1cm}
D(1,n)=
\begin{vmatrix}
 \mu_1 & \mu_2  & \mu_3 & \dots & \mu_{n+1} \\
 \mu_2 & \mu_3 & \mu_4 & \dots  & \vdots \\
\dots & \dots & \dots & \ddots & \vdots \\
\mu_{n+1} & & & & \mu_{2n+2}
\end{vmatrix} >0
\label{determinantal}
\ee

Given a Stieltjes series, there can be more than one function that has it as an asymptotic power series, since for instance there could be a second function differing purely by non-perturbative terms $e^{-\frac{1}{\lambda}}$. This potential complication is particularly relevant for the present work, so we need to have conditions on the Stieltjes series that ensure the uniqueness of the related Stieltjes function. There are criteria for uniqueness of a Stieltjes functions either in terms of the coefficients of a Stieltjes series, or in terms of the coefficients of a Stieltjes continued fraction.

We are not aware of any necessary and sufficient criterion for the uniqueness of a Stieltjes function in terms of the moments $\mu_n$. We present two criteria that are sufficient to prove uniqueness. The first criterion \cite{Bender:1999box, simon} states that if if $\{ \mu_k \}$ satisfy the determinantal conditions (\ref{determinantal}) and there exist constants C,M such that
\be
\mu_n \leq C M^n \, (2n)! 
\label{uniquecrit}
\ee
then the Stieltjes moment problem has a unique solution. The second criterion \cite{Bender:1999box, simon} is known as Carleman's criterion and it states that if $\{ \mu_k \}$ satisfy the determinantal conditions (\ref{determinantal}) and also the condition that
\be
\sum_{k=1}^\infty (\mu_k)^{-\frac{1}{2k}}=\infty
\label{carleman}
\ee
then the Stieltjes moment problem has a unique solution. Note that if $\mu_k$ satisfy the first criterion (\ref{uniquecrit}), they satisfy the second one (\ref{carleman}).

There are also criteria for the existence and uniqueness of a Stieltjes function in terms of the coefficients of a continued fraction \cite{baker, Bender:1999box}.

Finally if $f(z)$ is a power series that satisfies the determinantal conditions and Carleman's criterion, all sequences of $[M+J,M]$ Pad\'e approximants to $f(z)$ with $J\geq -1$ converge to its unique Stieltjes function in the cut plane \cite{baker, Bender:1999box, simon}.

\section{A bound for $\frac{K_{3/4}(x)}{K_{1/4}(x)}$.}
\label{appsegura}
In section \ref{analsection}, in order to demonstrate that $\langle \phi^4 \rangle$ is a Stieltjes function, we needed to argue that the following inequality
\be
(x^2+2x+\frac{1}{4})K_{\frac{1}{4}}(x)^2-\frac{3x}{2} K_{\frac{3}{4}}(x) K_{\frac{1}{4}}(x)-x^2 K_{\frac{3}{4}}(x)^2>0
\ee
holds for $x>0$. In this appendix we present a proof due to Javier Segura. The inequality is equivalent to
\be
\frac{K_{\frac{3}{4}}(x)}{K_{\frac{1}{4}}(x)} < \frac{\sqrt{x^2+2x+\frac{13}{16}}-\frac{3}{4}}{x} \equiv F(x)
\ee
and while there is literature on bounds for ratios of modified Bessel functions, {\it e.g.} \cite{segura} and references therein, it doesn't seem immediate to derive the bound above from those references. To prove the inequality, a possibility is to write the modified Bessel functions in terms of parabolic cylinder functions, undoing the change made in section \ref{analsection},
\be
\frac{K_{\frac{3}{4}}(x)}{K_{\frac{1}{4}}(x)}=1+\frac{U_1(2\sqrt{x})}{2\sqrt{x} U_0(2\sqrt{x})}
\ee
and then use the bound of Th. 3.2 in \cite{segura2},
\be
\frac{U_0(z)}{U_1(z)}>\frac{7z+3\sqrt{z^2+10}}{10}
\ee
This gives the bound
\be
\frac{K_{\frac{3}{4}}(x)}{K_{\frac{1}{4}}(x)}< 1+ \frac{5}{14x+3\sqrt{4x^2+10x}} \equiv B(x)
\ee
It remains to be argued that $B(x)<F(x)$, for $x>0$. For $x$ large enough,
\be
B(x)-F(x)=-\frac{33}{1024}\frac{1}{x^4}+{\cal O}(x^{-5})
\ee
which shows that $B(x)<F(x)$ for $x$ large enough. To conclude the proof, we note that $B(x)=F(x)$ has no real positive roots, which can be checked after some elementary algebra. Thus, $B(x)<F(x)$ along the whole positive real axis.

\section{ A second proof of $\langle \phi^2 \rangle_{\text{SD}}=\langle \phi^2 \rangle_{\text{Pad\'e}}$}
\label{appa}
A basic result in the theory of Pad\'e approximants is that both the numerators and denominators of Pad\'e approximants satisfy a set of three-term recursive relations, known as the Frobenius identities \cite{baker}. If we denote by $S$ either the numerator or the denominator of the Pad\'e approximant, the two Frobenius identities relevant for this proof are
\be
C(L-1,M) S^{[L,M]}(\lambda)=C(L,M) S^{[L-1,M]}(\lambda)-\lambda C(L,M+1) S^{[L-1,M-1]}(\lambda)
\label{frobenius}
\ee
and
\be
C(L,M-1) S^{[L,M]}(\lambda)=C(L,M) S^{[L,M-1]}(\lambda)-\lambda C(L+1,M) S^{[L-1,M-1]}(\lambda)
\label{frobenius2}
\ee
where C(L,M) are determinants of M $\times $ M matrices whose entries are coefficients of the power series (\ref{powerseries}),
\be
C(L,M)=
\begin{vmatrix}
c_{L-M+1} & c_{L-M+2} &\dots & c_{L} \\
c_{L-M+2} & c_{L-M+3} & \dots & c_{L+1} \\
\dots & \dots & \dots & \dots \\
c_L & c_{L+1} & \dots & c_{L+M-1} 
\end{vmatrix}
\ee
C(L,M) are determinants of Hankel matrices, something that we will soon put to use. We will now argue that this recursive relation (\ref{frobenius}) is equivalent to the ones derived from the Schwinger-Dyson, eq (\ref{recurp}). First, the perturbative expansion of $\langle \phi^2 \rangle_{\text{pert}}$ is
\be
\langle \phi^2 \rangle_{\text{weak}}=\sum_{k=0}^\infty c_k \lambda^k=1-3\lambda+24 \lambda^2-297 \lambda^3+4896 \lambda^4+\dots
\ee
with $c_0=1,c_1=-3$ and $c_n=-4nc_{n-1}-\sum_{k=1}^{n-2} c_k c_{n-k-1}$ for $n\geq 2$ \cite{oeis}. Moreover the Stieltjes continued fraction of this power series is known \cite{oeis}, 
\be
\langle \phi^2 \rangle_{\text{weak}}=
\cfrac{1}{1+\cfrac{3\lambda}{1+\cfrac{5 \lambda}{1+\cfrac{7\lambda}{1+\dots}}}}
\ee
and the $n$-th coefficient is $2n+1$. From the knowledge of the Stieltjes continued fraction we immediately obtain its Jacobi continued fraction

\be
\sum_{k=0} c_k \lambda^k = \cfrac{c_0}{1+a_0 \lambda - \cfrac{ b_1 \lambda^2}{1+a_1\lambda-\cfrac{b_2 \lambda^2}{1+a_2 \lambda-\dots}}}
\ee
with $a_0=3$, $a_n=4n+8$ for $n>0$ and $b_n=(4n-1)(4n+1)$. Lastly, we use an old result by Heilermann \cite{heilermann} (see Theorem 11 in \cite{advanced}) relating determinants of Hankel matrices to Jacobi continued fractions. Namely, if a series admits a Jacobi continued fraction presentation,
\be
\sum_{k=0}^\infty \mu_k x^k = \cfrac{\mu_0}{1+a_0x-\cfrac{b_1 x^2}{1+a_1x-\cfrac{b_2x^2}{1+a_2x-\dots}}}
\ee
then 
\be
\begin{vmatrix}
\mu_0 & \mu_1 & \dots & \mu_{n-1} \\
\mu_1 & \mu_2 & \dots & \mu_n \\
\vdots & \vdots & \ddots & \vdots \\
\mu_{n-1} & \mu_n & \dots & \mu_{2n-2}
\end{vmatrix}
= \mu_0^n b_1^{n-1} b_2^{n-2} \cdots b_{n-2}^2 b_{n-1}.
\ee
Applying this theorem to the two relevant determinants we obtain
\be
C(N,N)=(-3)^N \prod_{k=1}^N [(4k+1)(4k+3)]^{N-k}
\ee
and
\be
C(N,N+1)=\prod_{k=1}^N[(4k-1)(4k+1)]^{N-k+1}
\ee
so the Frobenius relations read
\be
S^{[N,N+1]}= \prod_{k=1}^N(4k+1) \left[(-1)^N S^{[N,N]}+\lambda \prod_{k=0}^N(4k+3) S^{[N-1,N]} \right]
\ee
and
\be
S^{[N,N]}=\prod_{k=1}^N(4k-1) \left[(-1)^N S^{[N-1,N]}-\lambda \prod_{k=1}^N(4k+1) S^{[N-1,N-1]} \right]
\ee
Now define $p_n,q_n$ by
\be
P^{[N,N]}=(-1)^N \prod_{k=1}^N (4k-1)^{N+1-k} \prod_{k=1}^{N-1} (4k+1)^{N-k} p_{2N}
\ee
\be
Q^{[N,N]}=(-1)^N \prod_{k=1}^N (4k-1)^{N+1-k} \prod_{k=1}^{N-1} (4k+1)^{N-k} q_{2N+1}
\ee
\be
P^{[N,N+1]}=\prod_{k=1}^N (4k-1)^{N+1-k} \prod_{k=1}^{N} (4k+1)^{N+1-k} p_{2N+1}
\ee
\be
Q^{[N,N+1]}=\prod_{k=1}^N (4k-1)^{N+1-k} \prod_{k=1}^{N} (4k+1)^{N+1-k} q_{2N+2}
\ee
Notice that
\be
\frac{P^{[N,N]}}{Q^{[N,N]}}=\frac{p_{2N}}{q_{2N+1}}, \hspace{1cm} 
\frac{P^{[N,N+1]}}{Q^{[N,N+1]}}=\frac{p_{2N+1}}{q_{2N+2}}
\ee
so we can take $\frac{p_{2N}}{q_{2N+1}}$, $\frac{p_{2N+1}}{q_{2N+2}}$ as the Pad\'e approximants. If we now plug these definitions into the Frobenius definitions, we find tha the $p_n,q_n$ polynomials satisfy the SD recursive relations, eq. (\ref{recurp}), thus proving that  $\langle \phi^2\rangle_{\text{SD}}= \langle \phi^2\rangle_{\text{Pad\'e}}$.

\section{A third proof of convergence of SD approximants}
\label{appconve}
In this appendix we provide a third proof of convergence of $\langle \phi^{2k} \rangle^n_{\text{SD}}$. While more explicit than the two demonstrations in the main text, this proof does not give any information about the limit function, beyond its existence. The starting point is the identity
\be
p_n^{(k)}q_{n-1}-p_{n-1}^{(k)}q_n =(2n-3)(-\lambda)\left(p^{(k)}_{n-1}q_{n-2}-p^{(k)}_{n-2}q_{n-1}\right)
\ee
valid for $n>k+1$. By iterating it, we learn that the difference between two consecutive approximants is
\be
\langle \phi^{2k}\rangle_{\text{SD}}^{n}-\langle \phi^{2k}\rangle_{\text{SD}}^{n-1}
=
(2n-3)!! (-\lambda)^{n-k-1} \frac{q_{k}(\lambda)}{q_{n}(\lambda) q_{n-1}(\lambda)}
\label{differsd}
\ee
We now want to prove that as the size $n$ of the truncations tends to infinity, this difference tends to zero, for any fixed $k$ and any $\lambda \geq 0$. Since in any pair of consecutive approximants, one belongs to the even subsequence and one to the odd one, the vanishing of their difference in the $n\to \infty$ limit implies that the two limiting functions are the same. The strategy we will pursue consists of the following steps:
\begin{enumerate}
\item{We note that  for $\lambda \geq 0$, $q_{k}(\lambda) \leq q_{k}(\lambda) q_{k+1}(\lambda)$. Thus
\be
\left| \langle \phi^{2k}\rangle_{\text{SD}}^{n}-\langle \phi^{2k}\rangle_{\text{SD}}^{n-1} \right| =
(2n-3)!! \lambda^{n-k-1} \frac{q_{k}(\lambda)}{q_{n}(\lambda) q_{n-1}(\lambda)} \leq
 (2n-3)!! \lambda^{n-k-1} \frac{q_{k+1}(\lambda)q_{k}(\lambda) }{q_{n}(\lambda) q_{n-1}(\lambda)} 
\label{qkbound}
\ee
}
\item{We will first prove that the RHS of (\ref{qkbound}) is a monotonic increasing function, so it reaches its maximum as $\lambda \rightarrow \infty$,
\be
\left| \langle \phi^{2k}\rangle_{\text{SD}}^{n}-\langle \phi^{2k}\rangle_{\text{SD}}^{n-1} \right| 
\leq 
\lim_{\lambda \rightarrow \infty} (2n-3)!! \, \lambda^{n-k-1} \frac{q_{k+1}(\lambda)q_{k}(\lambda) }{q_{n}(\lambda) q_{n-1}(\lambda)}
\ee
}
\item{Then we will conclude the proof by arguing that
\be
\lim_{n\rightarrow \infty} \lim_{\lambda \rightarrow \infty} (2n-3)!! \, \lambda^{n-k-1} \frac{q_{k+1}(\lambda)q_{k}(\lambda) }{q_{n}(\lambda) q_{n-1}(\lambda)}=0
\label{thelimit}
\ee
}
\end{enumerate}

The proof that the RHS of (\ref{qkbound}) is a monotonic increasing function in $\lambda$ for $\lambda>0$ is accomplished by showing that its derivative is positive for $\lambda>0$. First note that
\be
\lambda^{n-k-1} \frac{q_{k+1} q_k}{q_n q_{n-1}}= \lambda^{n-k-1} \frac{q_{n-2}q_{n-3}\dots q_{k+1} q_k}{q_n q_{n-1} q_{n-2}\dots q_{k+2}}= \frac{\lambda q_{n-2}}{q_n} \frac{\lambda q_{n-3}}{q_{n-1}}\dots \frac{\lambda q_{k}}{q_{k+2}}
\ee
and since the product of monotonic functions is monotonic, we only need to prove that
\be
\frac{\lambda q_m(\lambda)}{q_{m+2}(\lambda)}
\ee
is a monotonically increasing function for $\lambda \geq 0$. Here is where we take advantage of the fact that the denominators of all SD approximants are the same. Although $\langle \phi^{2k}\rangle_{\text{SD}}^{n}$ are not Pad\'e approximants for $k>2$, they share the denominators $q_n$ with $\langle \phi^{2}\rangle_{\text{SD}}^{n}$, so we can use properties of denominators of Pad\'e approximants of a Stieltjes series. In particular, by theorem 5.2.1 of \cite{baker}, all roots of $q_m$ are real and negative, and interlace with roots of $q_{m+2}$. Namely, $\lambda q_m$ and $q_{m+2}$ are polynomials of the same degree, call it $r$ and
\be
\frac{\lambda q_m(\lambda)}{q_{m+2}(\lambda)}= \frac{\lambda (\lambda+a_2)\dots (\lambda+a_r)}{(\lambda+b_1)(\lambda+b_2)\dots (\lambda+b_r)}
\ee
with all roots negative (except for $a_1=0$) and interlacing,
\be
0=-a_1 > -b_1>-a_2>-b_2 \dots  -a_r>-b_r
\ee
Then the first derivative of this rational function can be written as
\be
\left(\frac{\lambda q_m(\lambda)}{q_{m+2}(\lambda)}\right)'=
\frac{\lambda q_m(\lambda)}{q_{m+2}(\lambda)}
\left [\frac{b_1-a_1}{(\lambda+a_1)(\lambda+b_1)}+\dots+\frac{b_r-a_r}{(\lambda+a_r)(\lambda+b_r)}\right ] >0
\ee
so indeed it is strictly increasing. This concludes step 2) of our proof. 

We now proceed to prove step 3) in the proof, eq. (\ref{thelimit}). Since $q_k(\lambda) q_{k-1}(\lambda)$ is a polynomial of degree $k-2$, the numerator and the denominator of (\ref{thelimit})  are both polynomials of the same degree $n-2$. Thus, the limit as $\lambda \rightarrow \infty$ is a finite number. To derive this number, we need the leading coefficient of $q_k(\lambda)$. From the recursive definition we learn that for $q_{2m}(\lambda)$ the coefficient of $\lambda^m$ is
\be
\frac{4^m}{2} \left(\frac{\Gamma[m+\frac{3}{4}]}{\Gamma[\frac{3}{4}]}-\frac{\Gamma[m+\frac{1}{4}]}{\Gamma[\frac{1}{4}]}\right)
\ee
while for $q_{2m+1}(\lambda)$ the leading coefficient is
\be
4^m \frac{\Gamma[m+\frac{3}{4}]}{\Gamma[\frac{3}{4}]}
\ee 
Given these explicit values, the limit (\ref{thelimit}) amounts to
\be
\lim_{m\to \infty} (4m-1)!! \frac{2 \Gamma[\frac{3}{4}]}{16^m \Gamma[m+\frac{3}{4}] \left(\frac{\Gamma[m+\frac{3}{4}]}{\Gamma[\frac{3}{4}]}-\frac{\Gamma[m+\frac{1}{4}]}{\Gamma[\frac{1}{4}]}\right)
}
\sim \lim_{m\to \infty} \frac{\sqrt{2} \, \, \Gamma[\frac{3}{4}]^2}{\pi} \frac{1}{\sqrt{m}} =0
\ee
This concludes the proof that the rational approximants $\langle \phi^{2k}\rangle_{\text{SD}}^n$ converge pointwise to a single funtion in the range $\lambda \in [0,+\infty)$, as $n\to \infty$.

\bibliographystyle{JHEP}

\end{document}